\documentclass[10pt]{iopart}

\usepackage{iopams}
\usepackage{cite}
\usepackage{array}
\usepackage{amssymb}
\usepackage[cp1251]{inputenc}
\usepackage[T2A]{fontenc}
\usepackage{graphicx}
\usepackage[dvipsnames]{color}
\usepackage{colortbl}
\usepackage{multirow}
\usepackage{pdfsync}
\usepackage{footnote}
\usepackage{threeparttable}
\usepackage[english]{babel}
\begin{document}
\newcolumntype{-}{@{}}
\newcolumntype{C}{>{\displaystyle}c}
\newcolumntype{L}{>{\displaystyle}l}
\newcolumntype{R}{>{\displaystyle}r}
\newcolumntype{d}[1]{D{,}{,}{#1}}
\newcolumntype{=}{@{\,}C@{\,}}
\newcommand{\red}{\textcolor{black}}
\renewcommand{\thefootnote}{\arabic{footnote}}

\newcommand{\colorred}[1]{\textcolor{red}{\unskip #1}}
\newcommand{\colorblue}[1]{\textcolor{blue}{\unskip #1}}
\newcommand{\colorgreen}[1]{\textcolor[named]{OliveGreen}{\unskip #1}}
\newcommand{\colorggreen}[1]{\textcolor[named]{OliveGreen}{\unskip #1}}
\newcommand{\colorviolet}[1]{\textcolor[named]{Violet}{\unskip #1}}

\renewcommand{\colorblue}{} 
\renewcommand{\colorgreen}{} 
\renewcommand{\colorggreen}{} 
\renewcommand{\colorviolet}{} 
\renewcommand{\colorred}[1]{} 

\newcommand{\CR}{\colorred}
\newcommand{\CB}{\colorblue}
\newcommand{\CG}{\colorgreen}
\newcommand{\CGG}{\colorggreen}
\newcommand{\CV}{\colorviolet}

\title{Three-photon generation by means of third-order spontaneous parametric down-conversion in bulk crystals}

\author{N.~A.~Borshchevskaya$^{1,2}$, K.~G.~Katamadze$^{1,2,3}$, S.~P.~Kulik$^{1,2}$ and M.~V.~Fedorov$^1$}

\address{\emph{$^1$ A.~M.~Prokhorov General Physics Institute, Russian Academy of Sciences, Moscow, Russia\\
                $^2$ Faculty of Physics,  M.~V.~Lomonosov Moscow State University, Moscow, Russia\\
                $^3$ Institute of Physics and Technology, Russian Academy of Sciences, Moscow, Nakhimovsky prospect, 34, Russia}}

\ead{borschxyz@gmail.com}      
\vspace{10pt}
\begin{indented}
\item[]May 2015
\end{indented}

\begin{abstract}
We investigate the third order spontaneous parametric down-conversion process in a nonlinear media with inversion centers.
\colorred{Particularly, the detailed analysis, including}\colorblue{Specifically, we analyze in details} the three-photon
differential count rate in \colorred{a }unit frequency and angular region\colorblue{s}, total count rate and measurement
time \colorred{estimations,
is given}for rutile and calcite crystals \colorred{due to their}\colorblue{which have} comparatively large cubic susceptibilities.
Special attention is
given \colorred{for the considering}\colorblue{to consideration} of limited frequency and angular detection ranges in order to
calculate experimentally available detection rate values.
\end{abstract}

\vspace{2pc}
\ioptwocol

\section{Introduction}


\colorred{The g}\colorblue{G}eneration of photon-number (Fock) states of light is one of the main tasks in\colorred{the} quantum optics. They are interesting not only from fundamental, but also from\colorred{ a} practical point\colorgreen{s} of view\colorred{s} \colorred{due to}\colorblue{because of} their necessity for\colorblue{ solving problems of} quantum communication\colorblue{s} and linear optical quantum computation\colorblue{s}\colorred{ tasks}. While problems of single-photon and bi\colorred{-}photon\colorblue{-}state\colorred{s} generation are well studied, the direct and non post-selective generation of higher-order Fock states is still an attractive challenge.

In \colorred{the current}\colorblue{this} work we consider the\colorblue{ problem of} three-photon state generation\colorred{ problem}. \colorred{It's n}\colorblue{N}on-classical properties\colorblue{ of such states} enable heralded emission of photon pairs \cite{Sliwa2002, Barz2010, Wagenknecht2010} as well as\colorblue{ preparation of} three-body entangled states (for example, Greenberger-Horne-Zeilinger (GHZ) states \cite{Zeilinger1992, Bouwmeester1999})\colorred{ preparation}.

There are several proposed solutions for\colorblue{ the problem of} three-photon generation\colorred{ problem} such as cascaded or postselective second-order nonlinear processes \cite{Keller1998, Hubel2010, Shalm2012, Yukawa2013, Hamel2014, Guerreiro2014,  Krapick2015} and\colorblue{ formation of} approximate photon triplets\colorred{ formed} by SPDC photon pairs together with an attenuated coherent state \cite{Rarity1999a}. All these approaches give relatively low photon generation rates (up to 45/minute \cite{Shalm2012}) and have a big contribution of low-photon-number impurities.

On the other hand, the most natural way to generate three-photon states is a third-order spontaneous parametric down-conversion (TOSPDC). Unlike the other techniques, it enables to generate a three-particle entanglement in\colorred{ a} continuous degree\colorblue{s} of freedom, such as energy and momentum. This problem was previously studied theoretically, but to the best of our knowledge no experimental results were reported for direct spontaneous\colorblue{ generation of} triplets\colorred{ generation} based on $\chi^{(3)}$. \colorred{Just a}\colorblue{Only} stimulated third-order parametric down-conversion was demonstrated by seeding triplet modes \cite{Douady2004, Gravier2008}.

There are two approaches for TOSPDC generation: in bulk crystals \cite{Keller1998, Chekhova2005, Bencheikh2007, Dot2012} or in optical fibers \cite{Richard2011, Tarnowski2011, Corona2011,  Corona2011a, Huang2013, Huang2013a}. The bulk crystals allow\colorred{s simply} satisfy\colorblue{ing simply}\colorred{ a}\colorblue{ the} phase-matching condition using different polarization modes, but a spatial multimode structure of three-photon light and a limited crystal length complicate a high photon conversion probability and an effective detection. One can increase the interaction length and decrease the spatial mode number by us\colorred{e}\colorblue{ing}\colorred{ of} optical fibers. In this case phase-matching condition can be realized while the pump and three-photon light propagate in different spatial modes or by us\colorred{e}\colorblue{ing}\colorred{ of}\colorblue{ the} quasi-phase-matching. But a small mode overlap ($\sim 10^{-3}$ \cite{Huang2013a})  and a high absorption coefficient for\colorred{ a}\colorblue{ the} pump (in a visible and especially in UV case) also limit the generation rate.

Our work presents a theoretical description of\colorblue{ the} third\colorblue{-}order parametric down-conversion in crystals. Special attention\colorred{ we pay}\colorblue{ will be paid} to\colorred{ those}\colorblue{ crystals}  with inversion centers, having zero $\chi^{(2)}$ and hence prohibiting all three-wave processes. That is of great importance especially in the case of triple generation with  seeding beams because three-wave processes are much more intense and may suppress\colorred{ triplets} generation of \colorblue{ triplets} as well as\colorred{ its}\colorblue{ their} detection.





The paper is organized as follows. In Sec.~2  we  evaluate the  three-photon count rate in unit\colorblue{ ranges} of frequency and transverse wave vector\colorred{s ranges} and the integral count rate over all detectable frequencies and transverse wave vectors of scattered photons in a collinear degenerate regime of generation for\colorblue{ the} type-I and type-II
phase-matching. In Sec.~3 we get\colorred{ estimations}\colorblue{ estimates} of the minimal measurement time\colorred{ essential to}\colorblue{ sufficient for} distinguish\colorblue{ing} signal triple coincidences from noise ones. Then in Sec.~4 \colorred{estimations are carried out}\colorblue{our estimates are specified} for\colorred{ the} two\colorred{ specific} nonlinear crystals with inversion centers: calcite and rutile. And finally in Sec.~5 we discuss obtained results.

\section{Calculation of photon count rate}
\colorblue{As the}\colorred{The} process of third-order SPDC (TOSPDC) is similar to two-photon SPDC,\colorred{ therefore} in this section we\colorred{ will} follow the approach developed by D.~N.~Klyshko for biphotons in \cite{Klyshko1988a},\colorred{ but  extend}\colorblue{ though somewhat extended for}\colorred{ it to} the case of triplets.

\colorred{The decay of }\colorblue{Let a }pump photon in the mode $\vec{k}_p, \omega_p$\colorred{ into}\colorblue{ be decaying for} three photons in\colorred{ the} modes
$\vec{k}_1, \omega_1$, $\vec{k}_2, \omega_2$ and $\vec{k}_3, \omega_3$ and\colorblue{ let a pump be a monochromatic plane-wave propagating along the $z$-axis. Under these assumptions the photon energy and transverse momentum are conserved and the non-conservation of the longitudinal momentum determines the phase mismatch $\Delta k_z$:}\colorred{  is accompanied by the energy and momentum conservation law with $\Delta \vec{k}_z$ accuracy in longitudinal direction (coincided with pump photon direction). Assuming a monochromatic plane-wave pump we obtain blue the following equations:}
\begin{eqnarray}\label{phasemathing}
        \omega_1+\omega_2+\omega_3-\omega_p=0,\\
        \vec{q}_1+\vec{q}_2+\vec{q}_3=0,\\
        k_{1z}+k_{2z}+k_{3z}-k_p=\Delta k_z,
\end{eqnarray}
where $\vec{q}_i$ denote perpendicular components of $\vec{k}_i$.

In the second order of the perturbation theory TOSPDC  is described by the Hamiltonian
\begin{eqnarray}\label{hamiltonian}
  \nonumber H=\frac{1}{2}\int\limits_V  d^{3}\vec{r} \sum\limits_{\vec{k}_1, \vec{k}_2, \vec{k}_3} \;  \chi^{(3)} c_{k_1} c_{k_2} c_{k_3} a_{k_1}^\dag a_{k_2}^\dag a_{k_3}^\dag E_p\times\\ \nonumber
  \exp\left[{i(\vec{k}_p-\vec{k}_1-\vec{k}_2-\vec{k}_3)\vec{r}-i(\omega_p-\omega_1-\omega_2-\omega_3)t}\right]\\
   \hphantom{H=\frac{1}{2}\int\limits_V  d^{3}\vec{r} \sum\limits_{\vec{k}_1, \vec{k}_2, \vec{k}_3} \;  \chi^{(3)} c_{k_1} c_{k_2} c_{k_3} a_{k_1}^\dag a_{k_2}^\dag}+ H.c.,
\end{eqnarray}
where $E_{p}$ is\colorred{ an}\colorblue{ the} amplitude of\colorred{ a}\colorblue{ the} pump\colorblue{ considered as a} classical monochromatic plain  wave,
 $V$ is\colorred{ an}\colorblue{ the} interaction volume,
\begin{equation}\label{c_k}
    c_k\approx i\sqrt{\frac{ 2\pi \hbar \omega_k }{v}},
\end{equation}
 $v$\colorred{ ---}\colorblue{ is the} quantization volume and $a^\dag_{ki}$\colorred{ ---}\colorblue{ are the} photon creation operator\colorblue{s}\colorred{ in}\colorblue{ for}\colorred{ the} mode\colorblue{s} $k_i$\colorgreen{  ($i=1,\ 2,\ 3,\ p$)}.

 \colorred{According to Fermi's golden rule t}\colorblue{T}he rate of transition\colorblue{s per}\colorred{ in a} unit spectral\colorred{ $d\omega_i$} and transverse\colorblue{-}wave\colorblue{-}vector\colorblue{ ranges of each photon $d\omega_i$ and} $d\vec{q}_i$\colorred{ range of each photon in a triplet has the form}\colorblue{ is determined by the Fermi Golden Rule}
\begin{eqnarray}\label{P123_final}
\begin{array}[b]{R}
    \multicolumn{1}{L}
    {R_{\omega_1 \vec{q}_1 \omega_2 \vec{q}_2 \omega_3 \vec{q}_3}=
   \Gamma \;l^2 W_p   \mbox{sinc}^2 \left(\frac{\Delta k_z l}{2}\right) \times}\\[12pt]
  \multicolumn{1}{R}
   {\hphantom{\frac{dP}{ \omega_2 \omega_2 d\omega_3 d\omega_3 \Omega_2 \Omega_1\Omega_1\Omega_1\Omega_1\Omega_1\Omega_1\Omega_1}}
   \delta^{(2)}(\vec{q}) \delta(\red{\Omega})},
\end{array}
\end{eqnarray}
where
\begin{eqnarray}\nonumber
\delta^{(2)}(\vec{q})=\delta(q_{1x}+q_{2x}+q_{3x})\delta(q_{1y}+q_{2y}+q_{3y}),\\ \nonumber
\red{\Omega_i=\omega_i-\omega_p/3,\quad \delta (\Omega)=\delta(\Omega_1+\Omega_2+\Omega_3)},\\
\Gamma=\hbar \left[\chi^{(3)}\right]^2 \omega_1 \omega_2 \omega_3 /\left(c^{4} (2\pi)^2 n_1 n_2 n_3 n_p \right),\label{gamma}
\end{eqnarray}
$n_i$ \colorred{ is}\colorgreen{ are}\colorred{ an index of refraction}\colorred{ the refractive index}\colorgreen{ the refractive} \CV{indices}\colorred{ in the mode $i$ ($i=1,2,3,p$)}, $W_p$ is\colorred{ a}\colorblue{ the} pump power and $c$\colorred{ ---}\colorblue{ is the} speed of light.

To calculate the total three-photon generation rate, we\colorred{ need}\colorblue{ have} to integrate\colorblue{ the differential rate of Eq. (\ref{P123_final})} over\colorred{ the whole} spectral and angular region\colorblue{s},\colorred{ defined}\colorblue{ restricted by features of the}\colorred{ a} detection scheme:
\begin{eqnarray}
\label{eq:RT}
R_T
\approx I \Gamma l^2 W_p \mbox{, where} 	
	\\ \nonumber
I\equiv\int \mbox{sinc}^2 \left(\frac{\Delta k_z l}{2}\right)\,\colorblue{\delta^{(2)}(\vec{q}) \delta(\Omega)}\times\\
  \hphantom{qqqqqqqqqqqqqqqqqqqqq} \times  d\omega_1\,  d\omega_2\,  d\omega_3\,  d\vec{q}_1\,  d\vec{q}_2\, d\vec{q}_3 \label{eq:I}
\end{eqnarray}
\colorblue{The above-mentioned restrictions of the integration ranges will be applied to variables of all three photons. This is the approach needed for description of experimentally measurable correlations of photons. In contrast, in earlier publications} \cite{Chekhova2005, Bencheikh2007}\colorblue{ filtering related to the features of detectors was assumed to be applied  only to variables of one photon of a triplet whereas for two other photons ranges of integration were taken unlimited, which is sufficient only for description of the single-photon count rate.}

\begin{figure*}[t]\label{fig:type_I}
  	\includegraphics[scale=0.83]{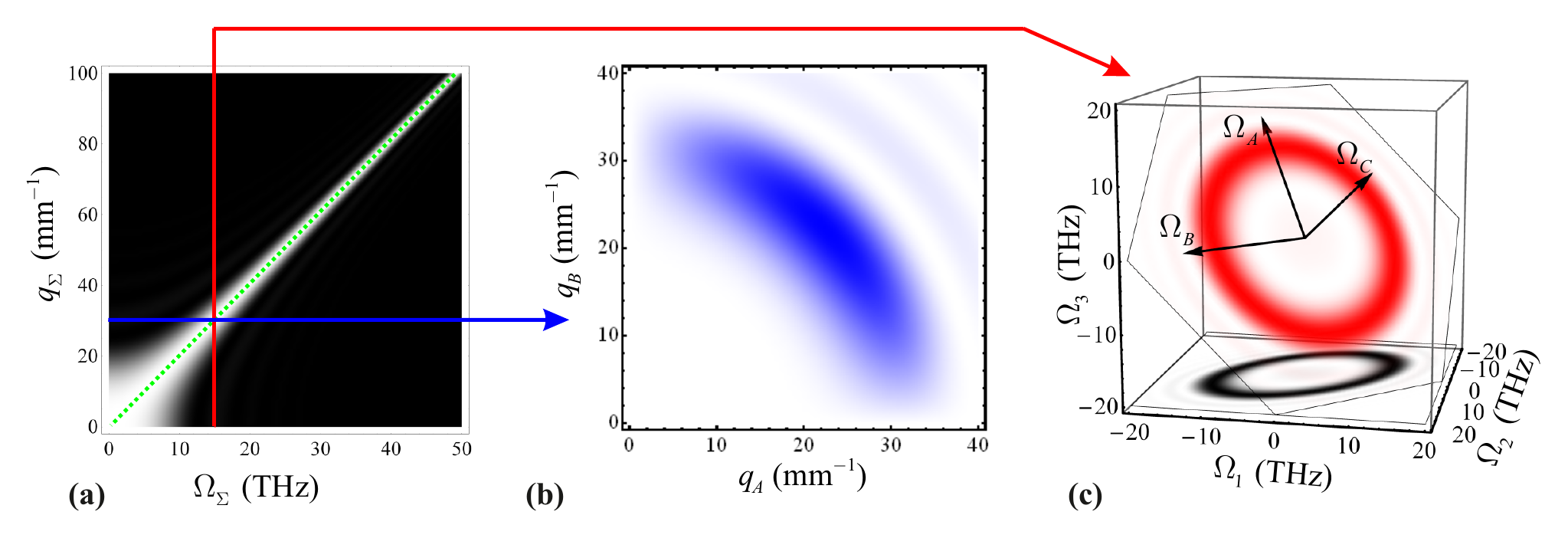}
  	\caption{(a) Function sinc$^2 \left[\Delta k_z(q_{\Sigma},\Omega_\Sigma) l/2\right]$ \colorgreen{and corresponding spectral distributions in $q_A-q_B$ parameters (b) and in $\Omega_A-\Omega_B$ parameters (c) for type-I TOSPDC. Green dotted line in (a) corresponds to the phase-matching curve defined by the equation $\Delta k_z(q_{\Sigma},\Omega_\Sigma)=0$.}  Calculation is made for 10~cm long rutile crystal at pump wavelength~532~nm. \colorred{The surface (a) is an analogue of the phase-matching surface of two-photon SPDC with the width equal to the square of blue and red areas (b,c).}}
\end{figure*}

In the biphoton case one can calculate the total count rate, integrating along the phase-matching curve, defined\colorred{ as}  \colorblue{ by the equation} $\Delta k_z(q_1, q_2=-q_1, \Omega_1, \Omega_2=-\Omega_1)=0$. This case  is comparatively simple as there are only\colorblue{ two independent integration variables}\colorred{ free parameters to integrate over}. \colorred{Next, }\colorblue{Below }we will derive the expressions for the curves\colorred{,} analogous to the biphoton phase-matching curve in the case of triplets for\colorblue{ the} type-I and type-II TOSPDC\colorred{ at} collinear degenerate regime of generation.

In order to do\colorred{ that}\colorblue{ this,} let us\colorred{ carry out the Taylor expansion of}\colorblue{ expand
$\Delta k_z$ in powers of $\Omega_i$, $q_{ix}$ and $q_{iy}$ up to the second order:}
\begin{eqnarray}\label{Delta_k_z_1}
\red{k_{z_{i}}=\sqrt{k_{i}^2-q_{i}^2},\quad k_{i}=n(\omega_{i})\frac{\omega_{i} }{c},\quad i=1,2,3}, \\
\nonumber\Delta k_z=k_{z1}+k_{z2}+k_{z3}-k_p=\left[k_1+k_2+k_3\right]_0-k_p+\\\hphantom{\Delta_z}
\nonumber\sum\limits_{i=1,2,3}\left[\frac{\partial k_i}{\partial q_{ix}}\right]_0 q_{ix}+\left[\frac{\partial k_i}{\partial q_{iy}}\right]_0 q_{iy}+\left[\frac{\partial k_i}{\partial \omega_i}\right]_0\Omega_i +\\ \hphantom{\Delta_z}
\frac{1}{2} \left[\frac{\partial^2 k_i}{\partial q_{ix}^2}\right]_0 q_{ix}^2
+\frac{1}{2}\left[\frac{\partial^2 k_i}{\partial q_{iy}^2}\right]_0 q_{iy}^2+
\frac{1}{2}\left[\frac{\partial^2 k_i}{\partial \omega_i^2}\right]_0\Omega_i^2 \label{Taylor}
\end{eqnarray}
Here
$[\ldots]_0$ means that\colorblue{ the} expression in brackets is evaluated at the exact collinear degenerate regime of generation.
  All\colorred{ the} mixed derivatives are equal\colorred{ to} zero.

Let's consider\colorblue{ separately}\colorred{ the} two types of phase-matching\colorblue{, type-I and type-II}.
\colorred{First of all,} \colorblue{
\subsection{Type-I phasematching}
In the case of}\colorred{ type-I}\colorblue{ phase matching of the type-I}
(e$\rightarrow$ooo)\colorred{. In this case all} the first-order derivatives\colorblue{ of $k_i$ in Eq. (\ref{Delta_k_z_1}) appear to be identical for different $i$}\colorred{ for different $k_i$  are equal}\colorblue{ and, as}\colorred{.  Taking
into account the phase-matching conditions:} $\sum \limits_i q_{ix,y}=0$ and $\sum \limits_i \Omega_i=0$,\colorred{ we get}\colorblue{ Eq. (\ref{Taylor}) takes the form}
\begin{eqnarray}\label{Delta_k_z_2}
\Delta k_z(q_{\Sigma},\Omega_\Sigma)=\beta\Omega_\Sigma^2 -\alpha q_{\Sigma}^2,
\end{eqnarray}
where
\begin{eqnarray*}
\alpha=-\frac{1}{2}\left[\frac{\partial^2 \Delta k_z}{\partial q_i^2}\right]_0 =\frac{3}{2}k_p, \\
\beta=\frac{1}{2}\left[\frac{\partial^2 \Delta k_z}{\partial \omega_i^2}\right]_0= \frac{1}{4\pi c^2} \left[\lambda^3  \frac{\partial^2 n_i}{\partial \lambda_i^2}\right]_0, \\
q_{\Sigma}^2=\sum \limits_{i=1,2,3} q_i^2, \quad \Omega_\Sigma ^2=\sum \limits_{i=1,2,3} \Omega_i^2.
\end{eqnarray*}
%

Thus, $\Delta k_z$ depends on two variables\colorblue{,} $q_{\Sigma}$ and $\Omega_{\Sigma}$\colorblue{,} and hence, the \CV{exact-}phase-matching curve \CR{can be} \CV{is defined by the equation}   $\beta\Omega_\Sigma^2 -\alpha q_{\Sigma}^2=0$ (\CR{see}\CV{the dotted green line in} Fig.~1a).

Now, let us\colorblue{ change}\colorred{ make several changes of variables in} the integration\colorblue{ variables in Eq.}~(\ref{eq:I})\colorred{ .
First, we change} $\Omega_1$, $\Omega_2$, $\Omega_3$\colorblue{ $\rightarrow$}\colorred{ to} $\Omega_A$, $\Omega_B$, $\Omega_C$
\CR{ (plotted by the green dotted line on Fig.~1c)} \CV{(Fig.~1c}):
\begin{eqnarray*}
\Omega_A=\frac{2\Omega_3-\Omega_1-\Omega_2}{\sqrt{6}},\quad
\Omega_B=\frac{\red{\Omega}_1+\red{\Omega}_2}{\sqrt{2}},\\
\Omega_C=\frac{\Omega_1+\Omega_2+\Omega_3}{\sqrt{3}}\colorred{.}
\end{eqnarray*}
\colorred{Then, we similarly change}\colorblue{ and}
$\vec{q}_1$, $\vec{q}_2$, $\vec{q}_3$\colorblue{ $\rightarrow$}\colorred{ to}
$\vec{q}_A$, $\vec{q}_B$, $\vec{q}_C$:

\begin{eqnarray*}
\vec{q}_{A}=\frac{2\vec{q}_3-\vec{q}_1-\vec{q}_2}{\sqrt{6}},\quad
\vec{q}_{B}=\frac{\vec{q}_1+\vec{q}_2}{\sqrt{2}},\\
\vec{q}_C=\frac{\vec{q}_1+\vec{q}_2+\vec{q}_3}{\sqrt{3}}.\\
\end{eqnarray*}
\colorred{Now }\colorblue{In these variables }$\delta(\Omega)\equiv \delta(\Omega_C)$ and $\delta^{(2)}(\vec{q})\equiv\delta^{(2)}(\vec{q}_C)$ in (\ref{eq:I}),\colorblue{ owing to which integrals over $\Omega_C$ and  $\vec{q}_C$ are easily taken.}\colorred{ and we need to integrate only over the rest variables in the planes  $\Omega_A$-$\Omega_B$ and ${q}_A$-${q}_B$.
Therefore, finally, we make the transformation  to }\colorblue{ As for the other variables, let us introduce the }polar coordinates\colorblue{ in the planes $(q_{Ax},q_{Ay})$, $(q_{Bx},q_{By})$, $(q_{A},q_{B})$, and $(\Omega_A,\Omega_B)$}\colorred{ $q_{\Sigma}$, $\Omega_{\Sigma}$, $\phi_A$, $\phi_B$,  $\phi_q$, $\phi_{\Omega}$}:
\begin{eqnarray*}
q_{Ax}=q_A\cos{\phi_A},\quad &q_{Ay}=q_A \sin{\phi_A},&\\
q_{Bx}=q_B\cos{\phi_B},\quad &q_{By}=q_B \sin{\phi_B},&\\
{q}_{A}=q_{\Sigma} \cos{\phi_q},\quad &q_{B}=q_{\Sigma} \sin{\phi_q},&\\
\Omega_A=\Omega_\Sigma \cos{\phi_\Omega},\quad &\Omega_B=\Omega_\Sigma \sin{\phi_\Omega}.
\end{eqnarray*}

\colorred{Now one can calculate the integral (\ref{eq:I}) in the }\colorblue{The Jacobian of transformation to }polar coordinates\colorred{ considering Jacobian}\colorblue{ is given by} $J=\Omega_\Sigma q_{\Sigma}^{3}\cos{\phi_q}\sin{\phi_q}$\colorblue{ and Eq. (\ref{eq:I}) takes the form}:
\begin{eqnarray}
 I= \nonumber \int dq_{\Sigma} d\Omega_\Sigma d\phi_q d\phi_A d\phi_B d\phi_{\Omega} \, J\,
  \mbox{sinc}^2 \left(\frac{\Delta k_z l}{2}\right)=\\
\label{IntType2}
\hphantom{I=}\frac{(2\pi)^3}{2}\int q_{\Sigma}^3 dq_{\Sigma} \Omega_\Sigma d\Omega_\Sigma \,\mbox{sinc}^2 \left(\frac{\Delta k_z l}{2}\right)
\end{eqnarray}

\CR{Each point  within t} \CV{T}he phase-matching area \CV{in the plane ($\Omega_\Sigma,q_{\Sigma}$)
is shown in white in} \CR{the $\{\Omega_\Sigma,q_\Sigma\}$
in the space on} Fig.~1a. \CV{Each point at the exact-phase-matching curve in this area}
corresponds to \colorred{a square of a fill blue ring quarter in the $q_A$-$q_B$ plane, which plays the role of the width of the phase-matching surface in the $q$-space} \CV{one quarter of a ring} \CR{-shaped} \CV{of the} \CG{spectral distribution in $q_A-q_B$ variables} (Fig.~1b),
and \CV{to} a\colorred{ square of a fill red ring in the  $\Omega_A$-$\Omega_B$ plane, which plays the role of the phase-matching surface in the $\Omega$-space}\colorgreen{ full} \CR{-}ring\CR{-shaped}
\CV{of the} distribution in \CV{the} $\Omega_A-\Omega_B$ variables (Fig.~1c).\colorgreen{The integrals}
\CR{of these} \CV{over the $q_A-q_B$ and $\Omega_A-\Omega_B$} distributions are included into the Jacobian $J$.

\CR{To find an approximate value of t} \CV{T}he integral \CV{over $q_\Sigma$ in} Eq. (\ref{IntType2}) \CV{ can be approximated by the product of the integrand at the exact-phase-matching curve with the width of the phase-matching area in the $q_\Sigma$ direction, $\Delta q_\Sigma$. The latter can be found from the equation} \CR{we substitute the integration over $q_\Sigma$ by the multiplication to the width of the phase-matching area
To find the width of the phase-matching\colorred{ curve} in $q_{\Sigma}-\Omega_{\Sigma}$-\colorred{space} with the  given and fixed $\Omega_\Sigma$ let us consider the corresponding $q_{\Sigma}$ (so that $\beta \Omega_\Sigma^2-\alpha q_{\Sigma}^2=0$)
and the mismatch $\Delta q_{\Sigma}$, satisfying the condition}:
\begin{eqnarray}\label{Delta_k_z_3}
(\beta \Omega_\Sigma^2-\alpha (q_{\Sigma}+\Delta q_{\Sigma})^2) l/2=\pi\; \Rightarrow\; q_{\Sigma} \, \Delta q_{\Sigma}=\frac{\pi}{l \alpha},
\end{eqnarray}
\CV{which reduces Eq. (\ref{IntType2}) to the form}
\colorred{From the phase-matching condition
\begin{equation*}
|\Delta k_z|=|\beta\Omega_\Sigma^2-\alpha q_{\Sigma}^2|\leq 2 \pi/l
\end{equation*}
one can extract the link between $q_{\Sigma}$ and $\Omega_\Sigma$ at perfect phase-matching $q_{\Sigma}=\Omega_\Sigma \sqrt{\beta/\alpha}$ and the width in $q_{\Sigma}$-direction $\Delta q_{\Sigma}=\pi/(\alpha l q_{\Sigma})$.
At perfect phase-matching $(\Delta k_z=0)$ one can also extract the relation $q_{\Sigma}=\Omega_\Sigma \sqrt{\beta/\alpha}$, and t,}
\begin{eqnarray}\label{P_new3}
\label{eq:IntI} I\colorgreen{\approx}\colorred{=} \frac{(2\pi)^3 }{2} \int \limits_0^{\Omega_{\Sigma max}}
\frac{\pi}{\alpha l}\frac{\beta}{\alpha}\Omega_\Sigma^3 d\Omega_\Sigma=\frac{ \pi^4 \beta}{\alpha^2 l}\Omega^4_{\Sigma max},\\ \nonumber
\mbox{where}\\ \nonumber
\Omega_{\Sigma max}\equiv
\min\left[
\frac{2\pi c}{\lambda_{min}}-\frac{\omega_p}{3},
\frac{\omega_p}{3}-\frac{2\pi c}{\lambda_{max}},
\frac{k_p\theta_{max}}{3} \sqrt{\frac{\alpha}{\beta}}
\right]
\end{eqnarray}
Here we have taken into account the limitation of the spectral (from $\lambda_{min}$ to $\lambda_{max}$) and angular (no more than $\theta_{max}$) detection range.

\colorblue{\subsection{Type-II phasematching} }
\colorred{Next, we }\colorblue{Let us }consider\colorblue{ now the TOSPDC process with the} type-II (e$\rightarrow$ooe) phase-matching\colorblue{. Let the}\colorred{ assuming an} optical axis\colorblue{ of a crystal is located}\colorred{ lying} in the\colorred{ $x-z$}\colorblue{ $(xz)$} plane. \colorblue{In this case }\colorred{Now }the first-order derivatives in\colorblue{ Eq.} (\ref{Taylor}) are not equal and\colorred{in this case $\Delta k_z$}\colorblue{ the} decomposition\colorblue{ of $\Delta k_z$ has}\colorred{ will have} the following form (we assume that the photons 1 and 2 are ordinary and the photon 3 is extraordinary):
\begin{eqnarray}\label{tailorII}
\nonumber\Delta k_z=\beta_o(\Omega_1+\Omega_2)+\beta_e\Omega_3 + \alpha_e q_{3x}+\\
\nonumber\red{\hphantom{qqqqqqqqqqqqqq}\frac{1}{2}\gamma_{ox}(q_{1x}^2+q_{2x}^2)+\frac{1}{2}\gamma_{ex}q^2_{3x}+}\\
\hphantom{qqqqqqqqqqqqqq}\frac{1}{2}\gamma_{oy}(q_{1y}^2+q_{2y}^2)+\frac{1}{2}\gamma_{ey}q^2_{3y},
\end{eqnarray}
where
\red{
\begin{eqnarray*}
\beta_{o}\equiv \frac{\partial k_{1,2}}{\partial \omega_{1,2}},\quad \beta_{e}\equiv \frac{\partial k_{3}}{\partial \omega_{3}},\quad \alpha_{e}\equiv \frac{\partial k_{3}}{\partial q_{3x}},\\
\gamma_{ox,y}\equiv \frac{\partial^2 k_{1,2}}{\partial^2 q_{1,2,x,y}},\quad \gamma_{ex,y}\equiv \frac{\partial^2 k_{3}}{\partial^2 q_{3,x,y}}.
\end{eqnarray*}
}
We\colorred{ have take}\colorblue{ took} into account\colorblue{ here} that inside a small angular range in\colorblue{ the} $x$-direction the terms\colorred{ including}\colorblue{ with} the second\colorblue{-order}  derivatives $\gamma_{o,e x}$ are much\colorred{ less}\colorblue{ smaller} than the terms\colorred{ including}\colorblue{ with} the first\colorblue{-order}  derivatives $\alpha_{o,e}$, but in the $y$-direction $\partial \Delta k_z/\partial q_{iy}$=0\colorblue{ and}, hence\colorblue{,} the second\colorblue{-order}  derivatives $\gamma_{o,e y}$\colorred{ can't be neglected}\colorblue{ have to be retained}.

\colorred{Let us denote }\colorblue{By denoting }$\gamma\simeq\gamma_{ey}\simeq\gamma_{oy}$,\colorred{ then}\colorblue{ we get}
\begin{eqnarray}\label{Delta k_z II}
\nonumber\Delta k_z=(\beta_o-\beta_e)(\Omega_1+\Omega_2)-\alpha_e (q_{1x}+q_{2x})+\\
\gamma(q_{1y}^2+q_{2y}^2),
\end{eqnarray}
\colorred{thus, }\colorblue{and the integral $I$ of Eq. (\ref{eq:I}) takes the form}
\begin{equation*}
I=\frac{2\pi}{l|\gamma|} \int \mbox{sinc}^2
\left(\frac{[\beta_+ \Omega_+ - \alpha_+ q_+] l}{2}\right)
d\Omega_+ d\Omega_- d q_{+}dq_{-},
\end{equation*}
where
\begin{eqnarray*}
\Omega_\pm =\frac{\Omega_1\pm\Omega_2}{\sqrt{2}},\quad q_{\pm}=\frac{q_{1x}\pm q_{2x}}{\sqrt{2}},\\
\alpha_+=\sqrt{2}\alpha_e,\quad \beta_+=\sqrt{2}(\beta_o-\beta_e),
\end{eqnarray*}
\colorblue{with the intervals of $q_{1,2y}^2$ where $\left| \Delta k_z l\right|<2 \pi$ estimated as $\Delta q_y^2={2\pi}/{l|\gamma|}$}\colorred{and  ${2\pi}/{l|\gamma|}\equiv \Delta q_y^2$ -- the range of $q_{1,2y}$, where $\left| \Delta k_z l\right|<2 \pi$}.

\begin{figure*}[t]\label{fig:type_II}
  	\includegraphics[scale=0.83]{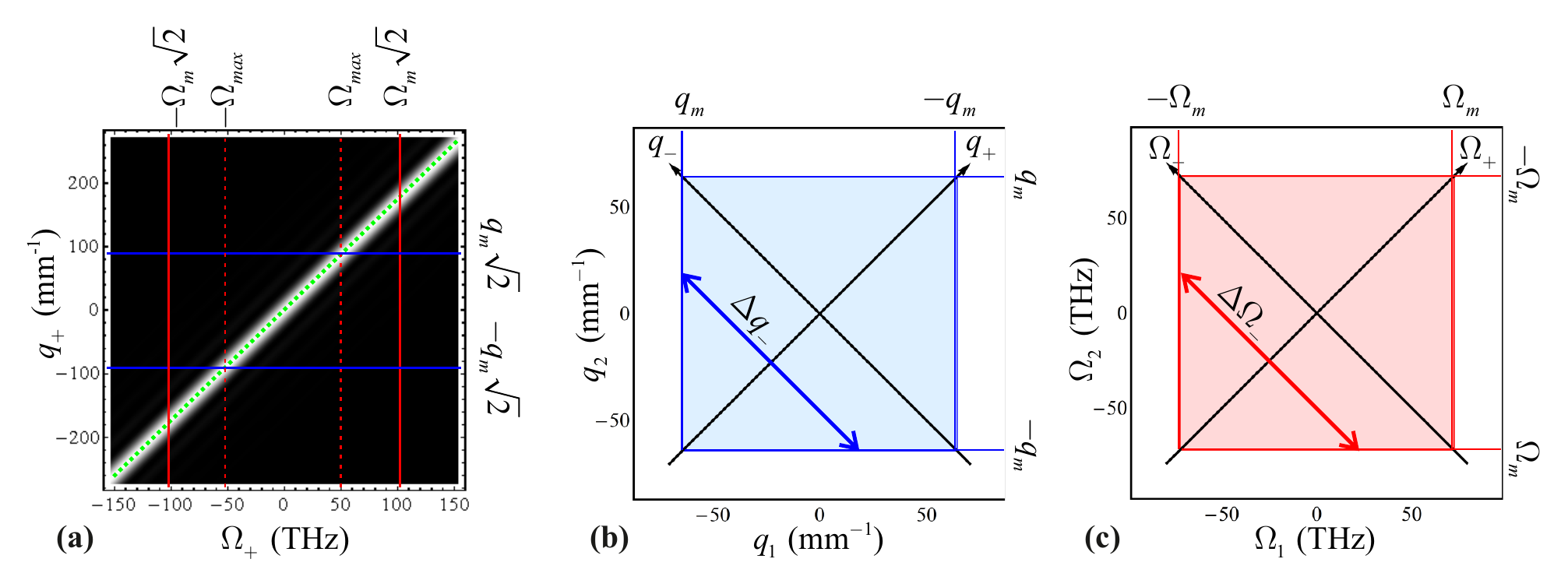}
  	\caption{(a) The function sinc$^2 \left[\CR{(\beta_+ \Omega_+ - \alpha_+ q_+)}\CG{\Delta k_z (\Omega_+, q_+)} l/2\right]$ and \CG{the integration areas over the variables $q_1$, $q_2$}  (b) and\CR{ in the $\Omega$-space}\CG{ over the variables $\Omega_1$, $\Omega_2$} (c) for type-II TOSPDC.\CG{ Green dotted line in (a) corresponds to the phase-matching curve defined by the equation $\Delta k_z (\Omega_+, q_+)=0$.} Calculation is made for a 1~mm long calcite crystal at the pump wavelength~532~nm.  \CR{The surface (a) is an analogue of phase-matching surface of two-photon SPDC with the width equal to the square of blue and red areas (b,c).}}
\end{figure*}

Hence, the \CV{exact-}phase-matching curve for type-II TOSPDC has the form $\beta_+ \Omega_+ - \alpha_+ q_+=0$ \CR{ plotted by} \CG{(the green dotted line} \CR{ on} \CV{in} {\CG Fig.~\ref{fig:type_II}a)}.
\CG{Each point of the phase-matching area $\{\Omega_+,q_+\}$ corresponds to the ranges $\Delta q_-$
(Fig.~\ref{fig:type_II}b) and $\Delta \Omega_-$ (Fig.~\ref{fig:type_II}c),}
%
%
\CR{The values $\Delta q_-$ and $\Delta \Omega_-$}\CG{which} can be found from phase-matching conditions:
\begin{eqnarray*}
\Delta q_-=2(q_{m}\sqrt{2}-|q_+|)=
2\left(q_{m}\sqrt{2}-\left|\frac{\beta_+}{\alpha_+}\Omega_{+}\right|\right)\\
\hphantom{qqqqqqqqqqq}\mbox {(shown as blue arrow on Fig.~{\ref{fig:type_II}b})},\\
\Delta \Omega_-=2({\Omega}_{m}\sqrt{2}-|\Omega_+|)\\
\hphantom{qqqqqqqqqqqq}\mbox {(shown as red arrow on Fig.~{\ref{fig:type_II}c})}\CR{,}\CG{.}\\
\end{eqnarray*}
\colorred{hence} \CV{Similarly to }\CR{As in} \CB{the type-I case we} \CR{substitute} \CV{approximate the integral} \CR{integration} over $q_+$ by the \CR{multipli-cation to} \CV{product of the integrand at the exact-phase-matching curve with} the width of the phase-matching area $\Delta q_+=4\pi/|\alpha_e|l$ and get \CR{the approximate value of the integral}
\begin{eqnarray}
\label{eq:IntII}
\nonumber I\CG{\approx}\CR{=}\frac{2\pi}{l |\gamma|}\int\limits_{-\sqrt{2}\Omega_{max}}^{\sqrt{2}\Omega_{max}} d\Omega_+ \cdot \Delta \Omega_-\Delta q_{+}\Delta q_{-}=\\
 \frac{32\pi^2}{ |\alpha_e\gamma |l^2} \int\limits_{-\sqrt{2}\Omega_{max}}^{\sqrt{2}\Omega_{max}} d\Omega_+ \cdot
\nonumber ({\Omega}_{m}\sqrt{2}-|\Omega_+|) \times\\ \hphantom{qqqqqqqqqqqqqqqq}\left(q_{m}\sqrt{2}-\left|\frac{\beta_+}{\alpha_+}\Omega_{+}\right|\right),
\end{eqnarray}
\colorred{Analogously as above Similarly to the case of the type-I phase-matching, } \CV{where} $\Omega_{max}$,  $\Omega_{m}$ and $q_{m}$ are determined by detection\colorblue{ ranges in} frequency and\colorblue{ angle}\colorred{ angular ranges}\colorred{ and correspond to each other at perfect phase-matching (see Fig.~\ref{fig:type_II}):}\CR{and} \colorgreen{shown in Fig.~2:}
\begin{eqnarray*}
q_{m}=k_p\theta_{max}, \quad
\Omega_{max}\equiv\min[{\Omega_m},\Omega_q],\quad \mbox{\CR{where} \CV{with}}\\
{\Omega_m}\equiv \min\left[\frac{2\pi c}{\lambda_{min}}-\frac{\omega_p}{3}, \frac{\omega_p}{3}-\frac{2\pi c}{\lambda_{max}}\right],\quad \Omega_q\equiv  q_{m} {\frac{\alpha_+}{\beta_+}}.
\end{eqnarray*}
\CR{Here $\Omega_m$ and $\Omega_q$ are limitations related to limits the range because of the spectral and angular filtering and $\Omega_q$ -- because of the angular filtering}.


\section{\colorblue{Evaluation of the }\colorred{Estimation of }measurement time}

The measurement time $T_3$ \colorblue{ can be defined}\colorred{ we will define} as the time\colorred{ essential to distinguish}\colorblue{ sufficient} for\CR{ distinguishing}\CG{ extracting the} signal triple coincidence\CR{s}\CG{ count rate} $R_s^{(3)}$ from noise\colorred{ ones} $R_n^{(3)}$. Mathematically\colorred{ it}\colorblue{ this} means that the
\CG{total number of triple coincidence counts} (\CR{measured as a difference between a sum of signal and noise counts and noise counts only}\CV{found as the difference of two measured rates, of the sum of signal and noise counts and, separately, of only  noise counts}\CG{) $N_s^{(3)}=R_s^{(3)}T_3$ is at least $t_{C,\infty}$ times bigger than its standard
deviation $\sigma_s^{(3)}$, where $t_{C,\infty}$ is the Student's $t$-factor for a confidence level $C$}\CR{:}\CGG{.
Taking into account the Poisson distribution of photo counts one can calculate the dispersions:
\begin{eqnarray*}
\sigma^{(3)}_{n}=\sqrt{R_n^{(3)}T_3},\quad \sigma^{(3)}_{n+s}=\sqrt{\left(R_n^{(3)}+R_s^{(3)}\right)T_3}\\
\sigma^{(3)}_{s}=\sqrt{\left(\sigma^{(3)}_{n}\right)^2+\left(\sigma^{(3)}_{n+s}\right)^2}=\sqrt{\left( 2 R_n^{(3)}+R_s^{(3)}\right)T_3}
\end{eqnarray*}
So, we obtain the equation for $T_3$:
}
%
%
\begin{eqnarray}\label{T1}
    \nonumber \CGG{N^{(3)}=}T_3 R_s^{(3)}=\CR{\zeta}\CG{t_{C,\infty}} \CR{\sqrt{\sigma_{n+s}}}\CG{\sigma_s^{(3)}}=\CR{\zeta}\CG{t_{C,\infty}} \sqrt{\left(2R_n^{(3)}+R_s^{(3)}\right)T_3}.
\end{eqnarray}
\colorblue{With g}\colorred{G}iven\colorred{ the} quantum efficiency $\eta$ and the noise count rate $R_n^{(1)}$ of each detector$^1$
\footnote
[0]
{
\noindent\rule{8cm}{0.4pt}
$^1$
We consider an ideal case when the\colorblue{ number of the} noise counts equals to \colorblue{ the number of} intrinsic detector's dark counts and\colorred{ all} the light noise is neglected.
} (we assume\colorblue{ that all detectors have approximately the same characteristics}\colorred{  for all detectors}), the temporal resolution of\colorred{ the} electronics (typically limited by a detector jitter) $\delta \tau$\colorblue{,} and $R_T$\colorred{,}  evaluated in (\ref{eq:RT}), we get
\begin{eqnarray}\label{T2a}
    R_n^{(3)}=\left(R_n^{(1)}\right)^3 \delta \tau ^2,\\
    R_s^{(3)}=\xi_3 R_T \eta^3,
\end{eqnarray}
where the \colorred{Here by adding the p}\colorblue{p}arameter $\xi_{3}$ (and also the parameter $\xi_2$ -- see below)
characterizes features of\CG{ non-polarized} beam splitters to be used in a possible experimental setup for dividing TOSPDC signal into three channels\colorred{ in an experimental set up that can be we have taken into account a potential experimental realization of coincidence schemes, the signal splits into several channels}.
Hence,
\begin{eqnarray}\label{T3b}
    T_3=\CR{\zeta^2}\CG{t_{C,\infty}^2} \frac{2\left(R_n^{(1)}\right)^3 \delta \tau ^2+\xi_3 R_T \eta^3 }{\left(\xi_3 R_T \eta^3 \right)^2}.
\end{eqnarray}
\colorblue{Similar }\colorred{Analogous }expressions can be derived for\colorred{ $T_2$ ($T_1$) ---} the minimal time\colorblue{ $T_2$ ($T_1$) required for distinguishing two-\colorred{ and one-}photon signal}\colorred{ to distinguish signal two-photon} coincidence\colorred{s (single)}\CG{ and single-photon} counts from\CG{ the}\colorblue{ noise}\colorred{ dark counts}:
\begin{eqnarray}\label{T2b}
    T_2=\CR{\zeta^2}\CG{t_{C,\infty}^2} \frac{2\left(R_n^{(1)}\right)^2 \delta \tau +\xi_2 R_T \eta^2 }{\left(\xi_2 R_T \eta^2 \right)^2},\\
    T_1=\CR{\zeta^2}\CG{t_{C,\infty}^2} \frac{ 2 R_n^{(1)} \delta \tau +R_T \eta }{\left(R_T \eta \right)^2}
\end{eqnarray}
and in case of \colorred{ the} two consistent\colorred{ beam splitters} 30/70 and 50/50\colorblue{ beam splitters}, which\colorblue{ provide}\colorred{ results in three channels with an} approximately equal power\colorblue{ in three channels, parameters $\xi_{3,2}$ are given by} $\xi_3=0.22$ and $\xi_2=0.75$.

\section{Example: calcite and rutile crystals}
\colorblue{Let us make estimates for two }\colorred{Now one can make estimations for the }specific crystals, calcite and rutile\colorblue{,}\colorred{ due to their two}\colorblue{ having} comparatively large cubic susceptibilities.\colorblue{ For these two crystals and for}\colorred{ For}  four pump wavelengths, 266, 325, 405 and 532~nm,\colorblue{ the results of calculations are presented  in Table \ref{table}}.  \CR{nd the both crystals w} \CV{By using data about crystal refractive indices of Refs. \cite{Ghosh1999,Gravier2006}, we found values of} \CR{the   We evaluated the
phase-matching angles the} angles between the optical axes of crystals \CR{the crystal} and the pump \CV{propagation} direction\CR{ relying on
refractive index values } \CV{providing collinear emission of TOSPDC photons. For these orientations of crystals, with the use of matrix elements determining $\chi^{(3)}$ \cite{Khadzhiiski1982,Penzkofer1983,Penzkofer1988,Borne2012}, and with  the dependence of $\chi^{(3)}$ on the angle between the pump propagation direction and the crystal optical axis \cite{midwinter1965} taken into account, we calculated values of the
effective cubic susceptibility $\chi^{(3)}_{eff}$ for both crystals and for all collinear TOSPDC
regimes indicated in Table \ref{table}. Together with  $\chi^{(3)}_{eff}$, we present in  Table~\ref{table} the total count rates (\ref{eq:RT}), calculated from  Eq. (\ref{eq:IntI}) for rutile and from Eq. (\ref{eq:IntII}) for calcite.}

\CR{,  accounting their dependence on evaluated phase-matching angles  and based on matrix elements of $\chi^{(3)}$ .
%
The values of $\chi^{(3)}_{eff}$, as well as , are shown in Table~\ref{table}.}
\definecolor{mygreen}{rgb}{0.7,1,0.7}
\definecolor{myred}{rgb}{1,0.7,0.7}
\definecolor{myyellow}{rgb}{1,1,0.7}
\begin{table*}[t]
\caption{
\CG{Calculated values of effective third-order nonlinear susceptibility $\chi^{(3)}_{eff}$, triplets generation rates $R_T$, measurement times necessary for registration of triple $T_3$ and double $T_2$ coincidences for different pump wavelengths $\lambda_p$ and power $W_p$, type of nonlinear media, its length $l$ and type of phase-matching, for different detectors and for the cases with presence $(+)$ and absence $(-)$ the cavity. Easy accessible in an experiment values marked as green, hardly accessible as yellow and unaccessible as red.
}}
\label{table}
\begin{tabular}[c]{c|c|c|c|c|c|c|c|c|c} \hline\hline
$\lambda_p$&Medium&$\chi^{(3)}_{eff}$&$W_p$,&Detector&l&Cavity&$R_T$&$T_3$&$T_2$\\
(nm)&&$(10^{-15}$esu)&(W)&&(mm)&&(Hz)&(days)&(days) \\  \hline
\multirow{2}*{266}&Calcite&\multirow{2}*{0.32}&\multirow{2}*{10}&\multirow{2}*{Si APD}&\multirow{2}*{0.1}&$-$&$4.0\cdot10^{-5}$&\cellcolor{myyellow}94&\cellcolor{myyellow}15\\  
&(e$\rightarrow$ooe)&&&&&$+$&$4.0\cdot10^{-2}$&\cellcolor{mygreen}$9.4\cdot10^{-2}$&\cellcolor{mygreen}$1.4\cdot 10^{-2}$\\ \hline
\multirow{2}*{325}&Calcite&\multirow{2}*{0.59}&\multirow{2}*{0.05}&Si APD&\multirow{2}*{0.1}&\multirow{2}*{$+$}&$1.1\cdot10^{-5}$&\cellcolor{myyellow}5 200&\cellcolor{myyellow}750\\ 
&(e$\rightarrow$ooe)&&&Super Cond.&&&$3.4\cdot10^{-6}$&\cellcolor{myyellow}18 000&\cellcolor{myyellow}1 000\\ \hline
\multirow{2}*{405}&Calcite&\multirow{2}*{0.76}&\multirow{2}*{0.5}&PMD&\multirow{2}*{0.1}&\multirow{2}*{$+$}&$1.8\cdot10^{-4}$&\cellcolor{myred}$8.1\cdot10^{10}$&\cellcolor{myred}$2.0\cdot10^{11}$\\ 
&(e$\rightarrow$ooe)&&&Super Cond.&&&$9.5\cdot10^{-6}$&\cellcolor{myyellow}6 200&\cellcolor{myyellow}$370$\\ \hline
\multirow{2}*{532}&Calcite&\multirow{2}*{0.88}&\multirow{2}*{10}&PMD&\multirow{2}*{0.1}&\multirow{2}*{$+$}&$1.8\cdot10^{-4}$&\cellcolor{myred}$8.2\cdot10^{10}$&\cellcolor{myred}$2.0\cdot10^{11}$\\ 
&(e$\rightarrow$ooe)&&&Super Cond.&&&$5.0\cdot10^{-6}$&\cellcolor{myyellow}1 200&\cellcolor{myyellow}$690$\\ \hline
\multirow{2}*{532}&Rutile&\multirow{2}*{71.6}&\multirow{2}*{10}&PMD&100&\multirow{2}*{$-$}&$1.5\cdot10^{-2}$&\cellcolor{myred}$1.1\cdot10^7$&\cellcolor{myred}$2.8\cdot10^7$\\ 
&(o$\rightarrow$eee)&&&Super Cond.&0.77&&$8.9\cdot10^{-7}$&\cellcolor{myred}$6.6\cdot 10^4$&\cellcolor{myred}$3.9\cdot 10^3$\\ \hline\hline
\end{tabular}
\end{table*}
Note that though rutile is a positive crystal and in the type-I phase-matching \CV{all three TOSPDC photons are not}
\CR{triple photons aren't} ordinary\CV{,} it can be shown that even in this case the calculation based on (\ref{eq:IntI}) results in inaccuracy of $\Delta k_z$ about 10 percents.

The phase-matching conditions in calcite at the mentioned pump wavelengths are satisfied for each type of phase-matching (e$\rightarrow$ooo, e$\rightarrow$ooe and e$\rightarrow$oee), but for all types except \CR{of} e$\rightarrow$ooe values of $\chi^{(3)}_{eff}$ are very small and, therefore, \CV{these cases are not included into } \CR{we don't include it in} Table~\ref{table}.

We paid \CR{a} special attention to \CV{calculations of the} optimal crystal length $l$ and angular detection range $\theta_{max}$ \CR{calculation}.

\colorblue{ Note, first, }\colorred{First of all, we should note, }
that the total count rate is \CR{directly} proportional to $l$ in
type-I \CR{case} (\ref{P123_final}), (\ref{P_new3}) and \CR{is} independent \CV{of} \CR{on} $l$
in type-II \CV{(\ref{P123_final}), (\ref{eq:IntII}) phase-matching cases}\CR{case}.
The last assertion is correct while in the phase-matching inequality (see (\ref{Delta k_z II}) and further)

\begin{equation*}
	|\Delta k_z|=\left|\alpha q +\frac{1}{2}\gamma q^2 \right|<\frac{2\pi}{l}
\end{equation*}
\CR{one can omit} the quadratic \CR{component} \CV{term can be omitted}. \CR{That} \CV{For calcite this}  is true for $l\gg l_{min}=4\pi\gamma/\alpha^2\sim0.05$~mm \CR{for calcite}.
But the crystal length can limit the detection angular range. For multi-mode detection (see Table~2)
we assume $\theta_{max}=\pi/2$$^2$
\footnote
[0]
{
\noindent\rule{8cm}{0.4pt}
$^2$
\CV{Of course,} \CR{We should own that} so big angles are outside \CV{of} the framework of our model but we suppose \CV{that, still, it remains reasonable} for rough estimates \CR{our calculations are suitable} .
} (we suppose, that all the SPDC radiation can be focused on the detector area). But for single-mode detection the angular range is defined by a gaussian mode divergence $\theta_{max}=\lambda/\pi w$, where $w$ \CR{--} is \CV{the} \CR{a} waist \CV{of the pump}, which should be less than \CR{a} \CV{the} spatial walk-off $\rho l$ (for the considered crystals $\rho\sim 0.1$). So, we obtain  $\theta_{max}\propto 1/l$. It means that in \CV{the case of} type-II \CV{phase-matching} \CR{case} we need to use as thin crystal as possible (we set $l=2l_{min}=0.1$~mm). \CV{In the case of} \CR{For a} type-I \CV{phase-matching} \CR{case} \CV{the crystal length $l$ has to be} \CR{we need to} decrease\CV{d until} the  \CR{while the} integration limits in Eq. (\ref{P_new3}) \CR{are} \CV{become} defined by the angular range. This means that the optimal crystal length is
\begin{equation*}
	l=
	\frac{k_p\lambda}{3\pi\rho} \sqrt{\frac{\alpha}{\beta}}\left/
	\min\left[
\frac{2\pi c}{\lambda_{min}}-\frac{\omega_p}{3},
\frac{\omega_p}{3}-\frac{2\pi c}{\lambda_{max}}
\right]\right..
\end{equation*}
Finally, for multi-mode detection and \CV{the} type-I \CV{phase-matching} \CR{case} \CV{the crystal has to be taken} \CR{we need to use} as long \CR{crystal} as possible (we set 100~mm).


\begin{table*}[t]
\label{tab:detectors}
  \centering
  \begin{threeparttable}[b]
\caption{Characteristics of single-photon detectors to be used for three-photons registration.}
\label{table_detectors}
\begin{tabular}[c]{c|ccccc} \hline\hline
\multirow{2}*{Type}&$\lambda_{min}$\--$\lambda_{max}$&Number&Quantum&Dark count rate $R_n^{(1)}$&Jitter $\delta\tau$\\
&(nm)&of spatial modes&efficiency $\eta$&(Hz)&(ps)\\ \hline
Si APD\tnote{1}&$\hphantom{1}400$--$1040$&Multi&0.1-0.7&100&350 \\ \hline
InGaAs APD\tnote{2}&$1000$--$1650$&Single&0.1&3000&200\\ \hline
Super Conductive\tnote{3}&$\hphantom{1}600$--$1700$&Single&0.2&1&50\\ \hline
PMD\tnote{4}&$\hphantom{1}950$--$1700$&Multi&0.01&50000&70\\ \hline\hline
\end{tabular}
 \begin{tablenotes}
    \item[1] Excelitas SPCM-AQRH-16 
    \item[2] IDQuantique ID210 
    \item[3] Scontel SSPD 
    \item[4] Hamamatsu R3809U-69 
  \end{tablenotes}
 \end{threeparttable}
\end{table*}



\section{Results and Discussion}
\CV{In calculations we took parameters of the most widely} \CR{We have considered the most commonly}
used \CR{types of} single-photon detectors (see Table~2)
and the most suitable cw lasers: DPSS lasers at 266 and 532~nm, diode blue-ray laser at 405~nm and HeCd gas laser at 325~nm. Typical powers are \CR{noted} \CV{given} in  Table~\ref{table}. It was shown\cite{Corona2011a} that \CV{the use of} \CR{using a} pulsed pump \CV{lasers} gives no \CR{benefit} \CV{advantages} \CR{to} \CV{for} TOSPDC generation.

\CR{We have a} \CV{A}lso \CR{used a chance to} \CV{we consider a possibility of increasing} \CR{increase} the pump power inside the crystal \CV{by means of mirror deposition at the rear and front faces of the crystal, which turns the latter into a cavity. Accordingly,  maximal growth of the pump intensity in a crystal (cavity) is characterized by the factor} $\varepsilon=$1000.\CR{times due to cavity mirrors deposition on the crystal surfaces}$^3$\footnote
[0]
{
\noindent\rule{8cm}{0pt}
$^3$
According to \cite{Siegman1986}, \CR{for a cavity with a mirrors reflectance} \CV{ the reflection coefficient of a lossless mirror is} $R_M\approx0.999$, which gives \CR{without losses, appearing after reflecting from the crystal surfaces,} $\varepsilon\approx (1-R_M)^{-1}\approx1000$. \CR{With nonzero l} \CV{L}osses \CV{in mirrors can decrease} \CR{at crystal surfaces the maximal }$\varepsilon$ \CV{making it} \CR{is} not higher than 10 for existing AR coatings.}
Unfortunately, \CV{this optimization cannot be used in} \CR{one can not use this option for} rutile crystal\CV{s} because of \CR{its} \CV{their} comparatively high adsorption.

\CV{In addition to } \CR{For all the cases we have calculated} the total triplet generation rate\CV{s} $R_T$\CV{, we have calculated and presented in  Table 1} \CR{and} the time, required for two- and three-photon coincidences\CV{,} (\ref{T3b}\CV{) and (}\ref{T2b}) \CV{at} \CR{for  $\zeta=3$}\CG{ $t_{0.998,\infty}=3$}. \CR{The best results are presented in Table~\ref{table}}.

 One can see, that one of the \CV{main} \CR{general} problem\CV{s} of TOSPDC detection is \CV{related to low}\CR{-} efficien\CV{cy} \CR{t} and high\CR{-} nois\CV{e} \CR{y} \CV{of} IR detectors. It is really difficult to extract a signal from \CR{a} noise even for triple coincidence measurements. So we \CV{found} \CR{find} only \CV{one combination of parameters when detection of} \CR{one} TOSPDC \CV{photons looks} \CR{detectable}\CGG{possible}. \CV{This is the} case \CV{of a calcite crystal with a mirror coverage deposited at rear and front faces, the crystal length $l=0.1$~mm, and the pump parameters $\lambda_p$=}\CR{: for }266~nm \CV{and $W_p$=}10~W\CV{. In this case we found} \CR{pump with a presence of the cavity: }$T_2=20$ and $T_3=135$~minutes\CR{ were calculated}.


\CGG{One more advantage of UV pump is the increase of the differential generation rate, because it is proportional to  $\omega_1\omega_2\omega_3\propto\omega_p^3$  (\ref{P123_final}).}

Another well-known problem is a low value\CR{s} \CV{of $\chi^{(3)}_{eff}$}. For rutile crystal $\chi^{(3)}_{eff}$\CR{ can be} \CR{increased in} \CV{is about} two orders \CV{higher if the crystal optical axis is taken parallel to the pump propagation direction. But for the phase-matching conditions to be satisfied one has to use a crystal with \CR{by means of using another angle between a pump and an optical axis, so the phase-matching conditions should be satisfied by means of} periodical\CR{ly} poling (quasi-phase-matching). \CV{This is an evident way for making the} \CR{In this case }type-I and type-II phase-matching conditions satisfied and TOSPDC photons detectable in visible range of wavelengths.} \CR{for visible (detectable by Si APD) three-photon generation can be realized.} \CV{Another possibility of increasing $\chi^{(3)}_{eff}$ is an addition of special impurities to crystals which would provide resonance enhancement of the third-order susceptibility}.  \CR{Also we should note, that nonlinear susceptibility highly increases near a resonance adsorption, so it can be enhanced by adding some special impurities inside a crystal lattice.}

\CV{One more problem is a really multi-mode structure of TOPDC generation in bulk crystals, which requires multi-mode detection of signals.
This problem can be solved by producing a wave-guide inside the crystal medium, as
proposed in \cite{Mazur2015}. In this way the length of interaction can be done arbitrary long.}

\section{Conclusion}
Finally, we have performed \CR{the} \CV{a} detailed analysis of three-photon generation in birefringent crystals with \CR{the} special attention \CV{paid} to calcite and rutile crystals. The analysis includes the calculation of differential generation rate in unit frequency and transverse  wave vectors range (\ref{P123_final}), total count rate (\ref{eq:RT}) for type-I and type-II phase-matching and measurement time,  required for distinguishing signal coincidences from noise \CR{ones} ((\ref{T2b}) and (\ref{T3b})).

The results show that the registration of TOSPDC in calcite is possible for the process with the pump at 266~nm with the presence of a cavity. All the other considered cases need\CR{s} too much time for three-photon registration.

This work was supported by \CV{the} Russian Science Foundation  (project 14-12-01338).

\section{\refname}
\renewcommand{\refname}{References}
\bibliographystyle{unsrt}
\bibliography{library}

\end{document}